\documentstyle[mprocl]{article}

\catcode`\@=11
\@addtoreset{equation}{section}
\catcode`\@=12
\begin{document}
\baselineskip8mm
\title{\vspace{-4cm}Simplest cosmological model with the scalar field
II. Influence of cosmological constant}
\author{A. Yu. Kamenshchik$^{1 \dag}$, \ I. M. Khalatnikov$^{1,2
\dagger}$ and \ A. V. Toporensky$^{3 *}$}
\date{}
\maketitle
\hspace{-6mm}
$^{1}${\em L. D. Landau Institute for Theoretical Physics,
Russian Academy of Sciences, Kosygin Street 2, Moscow, 117334,
Russia}\\
$^{2}${\em
Tel Aviv University, Raymond and Sackler
Faculty of Exact Sciences, School of Physics and Astronomy,
 Ramat Aviv, 69978, Israel}\\
$^{3}${\em Sternberg
Astronomical Institute, Moscow University, Moscow, 119899, Russia}\\
\\

Continuing the investigation of the simplest cosmological model with
the massive real scalar non-interacting inflaton field minimally
coupled to gravity we study an influence of the cosmological
constant on the behaviour of trajectories in
closed minisuperspace Friedmann-Robertson-Walker model.
The transition from chaotic to regular behaviour for large values of
cosmological constant is discussed.
Combining numerical calculations with qualitative analysis
both in configuration and phase space we present a convenient
classification of trajectories.
\\
$^{\dag}$ Electronic mail:  kamen@landau.ac.ru\\
$^{\dagger}$ Electronic mail:  khalat@landau.ac.ru\\
$^{*}$ Electronic mail: lesha@sai.msu.su\\

\section{Introduction}
\hspace{\parindent}
This paper continues investigation of our previous
paper $^{1}$ where we studied the dynamics of the cosmological model,
including gravity and minimally coupled scalar field with simple
potential including only massive term.  The dynamics of the
minisuperspace cosmological models with the massive real scalar field
for the flat, open and closed Friedmann universes was studied in
papers $^{2,3}$ in terms of phase space. It appeared that the
dynamics of closed model is more complicated than that of
open and flat models.  This dynamics allows the transitions from
expansion to contraction and the existence of points of maximal
expansion and minimal contraction in contrast with cases of open and
flat cosmologies.  Moreover, closed spherically symmetric models
cannot expand infinitely and should have the points of maximal
expansion provided the matter in the model under consideration
satisfies the condition of energodominance $^{4}$. The presence of
points of maximal expansion and minimal contraction opens the
possibility for the existence of the trajectories escaping
singularity and oscillating between turning points $^{5}$. The
possibility of existence of such trajectories or non-singular
universes filled with scalar field was discussed also earlier in Ref.
6. Recently such a dynamics was studed from the viewpoint of theory
of dynamical chaos $^{12}$.

Our previous work $^{1}$ was devoted to the
classification of trajectories in closed model and to the study the
idea about the fractal set of infinitely bouncing universes which was
put forward in Ref. 5.  Here we consider more rich model where
cosmological constant is included into consideration.  Dynamics in such
a model is rather peculiar and is of interest from mathematical point
of view. Besides, some attempts to reconstruct the potential of
inflaton field studying the anysotropy of relic cosmic background
radiation testify in favor of the existence of constant term in the
power expansion of this potential $^{7}$:  \[V(\varphi) = V_{0} +
\frac{m^{2} \varphi^{2}}{2} + \cdots ,\;\;V_{0} \neq 0.\] Apparently,
$V_{0}$ plays role of cosmological constant $\Lambda$.

Thus, we have an action
\begin{equation}
S = \int d^{4} x \sqrt{-g}\left\{\frac{m_{P}^{2}}{16\pi} (R -
2\Lambda) + \frac{1}{2} g^{\mu\nu}\partial_{\mu}\varphi
\partial_{\nu}\varphi -\frac{1}{2}m^{2}\varphi^{2}\right\},
\end{equation}
where $m_{P}$ is Planck mass.
The presence of cosmological constant $\Lambda$ makes possible
an existence of trajectories without points of maximal expansion,
i.e. the trajectories which begin and end in DeSitter regime
\begin{eqnarray}
&&a(t) \sim \exp(-Ht),\;\;t \rightarrow -\infty;\nonumber \\
&&a(t) \sim \exp(Ht), \;\;t \rightarrow \infty
\end{eqnarray}
or the trajectories which begin in singularity and end in DeSitter
expansion or vice versa. Here, in Eq. (1.2) $H$ denotes Hubble
constant
\[H = \sqrt{\frac{\Lambda}{3}}.\]

It is convenient to begin the consideration of possible classes of
trajectories with an investigation of the simplest
cosmological model when scalar field is massless, but cosmological
constant $\Lambda \neq 0$ (the model with $m = 0, \Lambda = 0$
was studied in some detail in Refs. 8,1,9). The second section of our
paper will be devoted to this case.  The class of trajectories
travelling between sigularities with possible multiple bounces will
be described in third section in terms of configuration space. In
fourth section we consider the trajectories of two types
DeSitter--singularity and DeSitter--DeSitter and the surface
separating them in phase space.

\section{Massless model with cosmological constant}
Equations of motion can
be now written down in the following form
\begin{equation}
\dot{y} =
-3 y z ,
\end{equation}
and
\begin{equation}
\dot{z} = z^{2}
+\frac{\Lambda}{3} -\frac{8\pi y^{2}}{3 m_{P}^{2}},
\end{equation}
where we have introduced notations
\[ y \equiv \dot{\varphi},\;\;z \equiv \frac{\dot{a}}{a}\]
and dots denote differentiation in respect with time.
First integral of motion looks as follows:
\begin{equation}
z^{2} + \frac{1}{a^{2}} = \frac{4\pi y^{2}}{3 m_{P}^{2}} +
\frac{\Lambda}{3}.
\end{equation}
As far as the variable $\varphi$ is not included into Eqs.
(2.3)--(2.5) the corresponding dynamical system can be represented by
the diagram of two-dimensional phase space with coordinates $(y,z)$,
see Fig. 1. Here two hyperbola
\begin{equation}
z = \pm \sqrt{\frac{4\pi y^{2}}{3 m_{P}^{2}} + \frac{\Lambda}{3}}
\end{equation}
separate the part of phase space accessible for the trajectories of
closed and open Friedmann - Robertson - Walker model (cf. Refs. 2,3).
There are four special points: points $A$ and $C$ are attracting and
repelling nodes respectively, while points $B$ and $D$ are saddle
ones. The coordinates of these points have the following values:
\begin{eqnarray}
&&y_{A} = 0,\;z_{A} = \sqrt{\frac{\Lambda}{3}};\;\;
y_{B} = \sqrt{\frac{m_{P}^{2} \Lambda}{8\pi}},\; z_{B} = 0;\nonumber
\\ &&y_{C} = 0,\;z_{C} = - \sqrt{\frac{\Lambda}{3}};\;\; y_{D} = -
\sqrt{\frac{m_{P}^{2} \Lambda}{8\pi}},\; z_{D} = 0.
\label{spec-point}
\end{eqnarray}
The region $I$ bounded by trajectories connecting the points $A$ and
$B$, $B$ and $C$, $C$ and $D$, $D$ and $A$ is covered by trajectories
which begins at point $C$, i.e since contraction according to
DeSitter exponential law at $t = - \infty$, then have a bounce or
in other words go through the point of minimal contraction
and end in the point $A$ corresponding to DeSitter expansion.

The region $II$ contains trajectories expanding from singularity
until DeSitter regime while the region $III$ contains trajectories
contracting from DeSitter point $C$ to singularity. Finally,
the region $IV$ contains trajectories which begin expansion from
singularity, then go through the points of maximal expansion and
then return to singularity again. The regions $II$ and $III$ are
separated from the region $IV$ by trajectories connecting
saddle points $B$ and $D$ with singularity.

It is interesting to notice that the picture analogous to that
described  above was obtained earlier for the hydrodynamical matter
in the presence of cosmological constant in Ref. 10. The diagram
similar to diagram in Fig.1 was constructed in the coordinates
$\varepsilon$ (energy) and $z$. The stationary unstable solutions
were obtained in Ref. 10 as well. Such a coincidence is quite natural
because massless scalar field corresponds to the hydrodynamical matter
with equation of state $\varepsilon = p$, where $p$ is pressure.

Inclusion of mass of scalar field makes the picture more
complicated.  Phase space becomes 3-dimensional and the opportunity of
multiple bounces or passages via points of maximal expansion arises.
Besides, the saddle points $B$ and $D$ disappear. Nevertheless, one can
speak roughly about three regimes again: they can be called
``singularity--singularity'', ``singularity--DeSitter'' and
``DeSitter--DeSitter''. First type of trajectories correspond to the
trajectories populating region $IV$ in massless case, but their
behaviour can be more involved due to passage through points of
minimal contraction and maximal expansion. Classification of these
trajectories in terms of configuration space $(a,\varphi)$ will be
given in third section of this paper. The second type of
trajectories corresponds to those in regions $II$ and $III$ in the
massless case while ``DeSitter-DeSitter'' trajectories correspond
to those from the region $I$. These two types of trajectories
will be studied in section 4.

\section{Travelling between singularities}
\hspace{\parindent}
It was shown in the preceding paper $^{1}$ that the points of maximal
expansion and those of minimal contraction can exist only in
Euclidean region. In the case of the presence of cosmological
constant $\Lambda$ the form of this region is given by equation
\begin{equation}
\varphi^{2} < \frac{3 m_{P}^{2}}{4 a^{2}} - \frac{\Lambda m_{P}^{2}}
{4 m^{2}}
\end{equation}
and this region is closed on the right at
\[a = \sqrt{\frac{3}{\Lambda}}.\]
The area where located the possible points of maximal expansion is
separated from that with points of minimal contraction by the curve
\begin{equation}
\varphi^{2} = \frac{m_{P}^{2}}{2\pi m^{2} a^{2}} -
\frac{\Lambda m_{P}^{2}}{4\pi m^{2}},
\label{separate}
\end{equation}
which crosses the horizontal axis at
\[a = \sqrt{\frac{2}{\Lambda}}.\]
Thus, geometry of this curve surrounding the region of possible
points of maximal contraction differs from that
consisting of two branches in the absence of $\Lambda$ (cf. Ref. 1).
The configuration of Euclidean region and the separating curve
(\ref{separate}) is shown in Fig. 2.

In the preceding paper $^{1}$ it was shown that the region of
possible points of maximal expansion has quite a regular structure.
In the left, closest to axis $a = 0$ part of this region there are
points of maximal expansion after which trajectory goes to
singularity. Then one can see the region where after the going
through the point of maximal expansion a trajectory undergoes the
``bounce'' i.e. goes through the point of minimal contraction. Then
we have the region where after going through the point of maximal
expansion trajectory has a ``$\varphi$ - turn '' i.e. has the
extremum in the value of the scalar field $\varphi$ and then fells
into singularity. Then one has the region corresponding to the
trajectories having bounce after one oscillation in $\varphi$ and
so on. Then studying the ``substructure'' of regions
corresponding to trajectories possessing bounces, one can see that
this substructure repeats the structure itself. In turn the
``subsubstrucure'' of regions having two bounces repeats the
general structure and so on and so forth. Continuing this process
{\it ad infinitum} we can get the set of infinitely bouncing
trajectories escaping the singularity. It is plausible to believe
that this set has a fractal nature $^{11}$ due to recurrent type of
the procedure of its construction $^{5,1,9,12}$.

The remarkable feature of the model with the cosmological constant
$\Lambda$ consists in the fact that this fractal structure of the
set of infinitely bouncing trajectories survives provided
the value of $\Lambda$ is small enough in comparison with the mass
of scalar field.

An important distinction between the case of small $\Lambda \neq 0$ and
$\Lambda = 0$  consists in the fact that an area of all the bouncing
intervals (i.e.  such locations of points of maximal expansion after
which the trajectories have bounce) is restricted from the right in
 configuration space by the new unstable periodical trajectory which
has no analogue in the case of $\Lambda = 0$.
 Trajectories situated on the left side of this periodical trajectory go
 to the regions of possible bounces while those situated on the right
 go to DeSitter point.Another distinction
is that substructure of all intervals contains subintervals,
corresponding to trajectories, going to DeSitter after bounce.

At some critical value of
\[\Lambda \sim 0.5 m^{2}\]
given by the numerical computation the
complicated structure of subregions of the part of Euclidean region
containing the points of maximal expansion vanishes.  Instead one has
only two regions (see Fig. 3): the left one containing the points of
maximal expansion after which trajectories fell to singularity and
the right one where after going through point of maximal expansion
and then via point of minimal contraction trajectories go to DeSitter
regime. Thus for the case of large values of $\Lambda$ the behaviour
of trajectories qualitatively reminds that for the case of massless
scalar field described in the preceding section.

It is interesting to notice that the transition from the chaotic
(fractal) structure of trajectories when we have infinite number of
subregions to the regular structure (Fig. 3) with two
subregions has jump-like character. There is not transitional
situation when one can see finite number of different subregions.

\section{Configuration of trajectories in the phase space}
\hspace{\parindent}
In this section we consider the relation between the trajectories
travelling between two DeSitter regimes and those which begin at
DeSitter point and end in the singularity or viceversa. The boundary
between these two classes of trajectories is rather complicated and
fractal-like two dimensional surface in three-dimensional phase
space. To describe it we shall use smooth two-dimensional
projections ignoring its fractal nature.

It is convenient to introduce the following coordinates in the phase
 space:
\begin{equation}
x = m_{P} \varphi,\;\; y = \frac{m_{P} \varphi}{m},\;\;
 z = m \frac{\dot{a}}{a}.
\label{norma}
\end{equation}

At $\Lambda \ne 0$ we have two De-Sitter points at
$ x=0 $ \\ $ y=0 $ \\ $ z= \pm m \sqrt{\Lambda/3} $
Some part of trajectories beginning in the neighborhood
of DeSitter points go to another DeSitter point while the rest
of trajectories fell to singularity or in terms of phase space
to an infinite sphere (see Refs. 2,3). To distinguish between
these two classes of trajectories we make numerical integration of
equations of motion starting from lower hemisphere around De-Sitter
point. For concreteness, the radius of this hemisphere is chosen as $ r
= z/3$. The behaviour of trajectories is shown in Fig. 4. It is
convenient to parametrize initial conditions for trajectories on
hemisphere by two angles $\phi$ and $\theta$ which are defined as
\[\tan(\phi)=y/x;\;\;\;\sin (\theta)=(\sqrt{\Lambda/3}+z)/r.\]
While $t \rightarrow \infty$ a trajectory may go to singularity or
to DeSitter asymptotic point. In Fig. 4 dashed area corresponds to
trajectories going to singularity. It is necessary to emphasizes that
this picture is a rough one - it exists a fine structure of
trajectories escaping sungularity inside the dashed area due to
 bounces. In Fig. 4b such a structure is remarkable.  For every value
  of relation $\Lambda/m^{2}$ there are two distinguished values of an
 angle $\theta$: $\theta_{min}$ such that at $\theta < \theta_{min}$
 all the trajectories go to singularity independently of value of angle
$\phi$ and $\theta_{max}$ such that all the trajectories go to the
second DeSitter point while $\theta > \theta_{max}$.

It is interesting to trace out the dependence of these angles on the
relation $\Lambda/m^{2}$. At small values of $\Lambda/m^{2}$
these angles almost coincide $\theta_{min} \approx \theta_{max}$.
This fact can be explained by the fact that the distance between
DeSitter points in the phase space is large in comparison with mass
parameter and the trajectories manage to have so many rotations
around axis $z$ that the dependence of their final destiny on the
initial angle $\phi$ disappear. It would be reasonable to expect
that in the limit $\Lambda/m^{2} \rightarrow \infty$ these two
limiting angles should also coincide giving the angle separating
different type of trajectories in massless case (cf. sec. 2), however
it is not the case because in our parametrization small values of
mass $m$ are compensated by possible large values of the field
$\varphi$ and one does not have smooth transition to the massless
model.  Therefore, it is interesting to carry out numerical
investigation of massless model considering since the beginning
two-dimensional phase space. In this case one has only one parameter
for the trajectories going from lower DeSitter point. It is
``latitude'' angle $\theta$. The value of $\theta$ separating
trajectories going to singularity in the limit $r \rightarrow
0 $ appears to be equal
\begin{equation}
 \theta_{numerical} \approx 1.2 .
 \label{numerical}
 \end{equation}
 One can easily estimate
theoretical value of this angle connecting DeSitter point with one of
two saddle points of two-dimensional phase diagram (see Fig. 1). Using
Eqs. (\ref{spec-point}) together with normalization given by Eqs.
(\ref{norma}) one can easily get
 \begin{equation} \theta_{theor} = \arctan \sqrt{\frac{8\pi}{3}}
\approx 1.3.
   \label{theor}
  \end{equation}
  Thus we have seen that the
results of theoretical consideration and numerical simulation are in a
 good agreement.

\section*{Acknowledgement}
 This work was supported by Russian Foundation for Basic Research
 via grants No 96-02-16220 and No 96-02-17591.

\begin{description}
\item[\rm 1.] A.Yu. Kamenshchik, I.M.
Khalatnikov and A.V. Toporensky, {\it Int. J. Mod. Phys.} {\bf D}, to
appear, // gr-qc/9801064.
\item[\rm 2.] V.A. Belinsky, L.P. Grishchuk, Ya.B.
Zel'dovich and I.M. Khalatnikov, {\it J. Exp. Theor. Phys.} {\bf 89},
346 (1985).
\item[\rm 3.] V.A. Belinsky and I.M. Khalatnikov, {\it
J.  Exp. Theor.  Phys.}  {\bf 93}, 784 (1987); V.A. Belinsky, H.
Ishihara, I.M. Khalatnikov and H. Sato, {\it Progr. Theor. Phys.}
{\bf 79}, 676 (1988).
\item[\rm 4.] G.A. Burnett, {\it Phys. Rev.}
{\bf D51}, 1621 (1995).
\item[\rm 5.] D.N. Page, {\it Class. Quantum
Grav.} {\bf 1}, 417 (1984).
\item[\rm 6.] L. Parker and S.A.
Fulling, {\it Phys. Rev.} {\bf D7}, 2357 (1973); A.A. Starobinsky,
{\it Pisma A.J.} {\bf 4} 155 (1978).
\item[\rm 7.] A. Melchiori, M.
Sazhin, V. Shulga and N. Vittorio, in preparation; V.N. Lukash and
E.V.  Micheeva, {\it Gravitation and Cosmology} {\bf 2} 247 (1996).
\item[\rm 8.] V.A.
Belinsky and I.M. Khalatnikov, {\it J.  Exp.  Theor.  Phys.}  {\bf
63}, 1121 (1972).
\item[\rm 9.] I.M. Khalatnikov and A.Yu.
Kamenshchik, {\it Phys.  Rep.}, to appear.
\item[\rm 10] E.S. Nikomarov and I.M. Khalatnikov, {\it ZhETF}
{\bf 75}, 1176 (1978); I.M. Khalatnikov and V.A. Belinsky,
{\it Qualitative Cosmology} in {\it Problems of modern experimental and
theoretical physics}, dedicated to 80th anniversary of Yu.B. Khariton,
ed. A.P. Alexandrov, Leningrad, 1984, p. 299 (in Russian).
\item[\rm 11.] B. Mandelbrot,
{\it The fractal geometry of nature}, (San Francisco, Freeman, 1982).
\item[\rm 12.] N.J.Cornish, E.P.S.Shellard "Chaos in quantum
cosmology" // gr-qc/9708046
 \end{description}
 \newpage

 {\bf Captions to Figures}\\
 {\bf Fig. 1} In Fig. 1 the phase diagram for the model with massless
scalar field is presented. Two hyperbolah going thgrough points $A$ and
$C$ separate the part of phase space accessible for the trajectories of
closed and open Friedmann - Robertson - Walker model.
Points $A$ and $C$ are attracting and
repelling nodes respectively, while points $B$ and $D$ are saddle
ones.
The region $I$ bounded by trajectories connecting the points $A$ and
$B$, $B$ and $C$, $C$ and $D$, $D$ and $A$ is covered by trajectories
which begins at point $C$, i.e since contraction according to
DeSitter exponential law, then have a bounce
and end in the point $A$ corresponding to DeSitter expansion.
The region $II$ contains trajectories expanding from singularity
until DeSitter regime while the region $III$ contains trajectories
contracting from DeSitter point $C$ to singularity. Finally,
the region $IV$ contains trajectories which begin expansion from
singularity, then go through the points of maximal expansion and
then return to singularity again. The regions $II$ and $III$ are
separated from the region $IV$ by trajectories connecting
saddle points $B$ and $D$ with singularity.\\
{\bf Fig. 2.} Solid line denotes the boundary between Lorentz and
Euclidean regions, dashed line corresponds to the curve separating
the points of minimal contraction and maximal expansion.\\
{\bf Fig. 3.}  Separating curve (solid) and boundary of bouncing
interval (dashed) in the case of large $\Lambda/m^2$. The bouncing
interval is located between these two curves. \\
{\bf Fig. 4} In Fig. 4 dashed area corresponds to the trajectories
going out DeSitter contracting point and falling to singularity.
The angles $\theta$ and $\varphi$ parametrize the direction of the
trajectory in the initial DeSitter point. White area corresponds to
trajectories going between two DeSitter points. In Fig. 4a
$\Lambda = 0.05 m^{2}$, in Fig. 4b $\Lambda = 0.5 m^{2}$, in Fig. 4c
$\Lambda = 50 m^{2}$.
\end{document}